\documentclass[twocolumn, amsfonts, amssymb, amsmath, showkeys, superscriptaddress]{revtex4-2}
\usepackage{amsmath,amssymb,amsfonts}
\usepackage[utf8]{inputenc}
\usepackage[T1]{fontenc}
\usepackage{algorithmic}
\usepackage{graphicx}
\usepackage{textcomp}
\usepackage{url}
\usepackage{ragged2e}
\usepackage{hyperref}

\begin{document}
\preprint{}
\title{Rapid Autotuning of a SiGe Quantum Dot into the Single-Electron Regime with Machine Learning and RF-Reflectometry FPGA-Based Measurements}

\author{Marc-Antoine Roux}
\email{email: Marc-Antoine.M.Roux@USherbrooke.ca}
\affiliation{Département de physique, Université de Sherbrooke, Sherbrooke (Québec) J1K 2R1, Canada}
\affiliation{Institut quantique, Université de Sherbrooke, Sherbrooke (Québec) J1K 2R1, Canada}
\author{Joffrey Rivard}
\affiliation{Département de physique, Université de Sherbrooke, Sherbrooke (Québec) J1K 2R1, Canada}
\affiliation{Institut quantique, Université de Sherbrooke, Sherbrooke (Québec) J1K 2R1, Canada}
\author{Victor Yon}
\affiliation{Institut quantique, Université de Sherbrooke, Sherbrooke (Québec) J1K 2R1, Canada}
\affiliation{Institut interdisciplinaire d'innovation technologique, Université de Sherbrooke, Sherbrooke (Québec) J1K 0A5, Canada}
\author{Alexis Morel}
\affiliation{Département de physique, Université de Sherbrooke, Sherbrooke (Québec) J1K 2R1, Canada}
\affiliation{Institut quantique, Université de Sherbrooke, Sherbrooke (Québec) J1K 2R1, Canada}
\author{Dominic Leclerc}
\affiliation{Département de physique, Université de Sherbrooke, Sherbrooke (Québec) J1K 2R1, Canada}
\affiliation{Institut quantique, Université de Sherbrooke, Sherbrooke (Québec) J1K 2R1, Canada}
\author{Claude Rohrbacher}
\affiliation{Département de physique, Université de Sherbrooke, Sherbrooke (Québec) J1K 2R1, Canada}
\affiliation{Institut quantique, Université de Sherbrooke, Sherbrooke (Québec) J1K 2R1, Canada}
\author{El Bachir Ndiaye}
\affiliation{Département de physique, Université de Sherbrooke, Sherbrooke (Québec) J1K 2R1, Canada}
\affiliation{Institut quantique, Université de Sherbrooke, Sherbrooke (Québec) J1K 2R1, Canada}
\author{Felice Francesco Tafuri}
\affiliation{Keysight Technologies, 1400 Fountaingrove Pkwy., Santa Rosa, California 95403, USA}
\author{Brendan Bono}
\affiliation{Keysight Technologies, 1400 Fountaingrove Pkwy., Santa Rosa, California 95403, USA}
\author{Stefan Kubicek}
\affiliation{IMEC, Kapeldreef 75, 3001 Leuven, Belgium}
\author{Roger Loo}
\affiliation{IMEC, Kapeldreef 75, 3001 Leuven, Belgium}
\affiliation{Department of Solid-State Sciences, Ghent University, Krijgslaan 281, building S1, 9000 Ghent, Belgium}
\author{Yosuke Shimura}
\affiliation{IMEC, Kapeldreef 75, 3001 Leuven, Belgium}
\author{Julien Jussot}
\affiliation{IMEC, Kapeldreef 75, 3001 Leuven, Belgium}
\author{Clément Godfrin}
\affiliation{IMEC, Kapeldreef 75, 3001 Leuven, Belgium}
\author{Danny Wan}
\affiliation{IMEC, Kapeldreef 75, 3001 Leuven, Belgium}
\author{Kristiaan De Greve}
\affiliation{IMEC, Kapeldreef 75, 3001 Leuven, Belgium}
\affiliation{Department of Electrical Engineering, KU Leuven, 3001 Leuven, Belgium}
\affiliation{Proximus Chair in Quantum Science and Technology, KU Leuven, 3001 Leuven, Belgium}
\author{Marc-André Tétrault}
\affiliation{Institut quantique, Université de Sherbrooke, Sherbrooke (Québec) J1K 2R1, Canada}
\affiliation{Institut interdisciplinaire d'innovation technologique, Université de Sherbrooke, Sherbrooke (Québec) J1K 0A5, Canada}
\author{Dominique Drouin}
\affiliation{Institut quantique, Université de Sherbrooke, Sherbrooke (Québec) J1K 2R1, Canada}
\affiliation{Institut interdisciplinaire d'innovation technologique, Université de Sherbrooke, Sherbrooke (Québec) J1K 0A5, Canada}
\author{Christian Lupien}
\affiliation{Département de physique, Université de Sherbrooke, Sherbrooke (Québec) J1K 2R1, Canada}
\affiliation{Institut quantique, Université de Sherbrooke, Sherbrooke (Québec) J1K 2R1, Canada}
\author{Michel Pioro-Ladrière}
\affiliation{Département de physique, Université de Sherbrooke, Sherbrooke (Québec) J1K 2R1, Canada}
\affiliation{Institut quantique, Université de Sherbrooke, Sherbrooke (Québec) J1K 2R1, Canada}
\author{Eva Dupont-Ferrier}
\affiliation{Département de physique, Université de Sherbrooke, Sherbrooke (Québec) J1K 2R1, Canada}
\affiliation{Institut quantique, Université de Sherbrooke, Sherbrooke (Québec) J1K 2R1, Canada}

\date{\today}

\begin{abstract}
This work has been submitted to the IEEE for possible publication. Copyright may be transferred without notice, after which this version may no longer be accessible.\\

Spin qubits need to operate within a very precise voltage space around charge state transitions to achieve high-fidelity gates. However, the stability diagrams that allow the identification of the desired charge states are long to acquire. Moreover, the voltage space to search for the desired charge state increases quickly with the number of qubits. Therefore, faster stability diagram acquisitions are needed to scale up a spin qubit quantum processor. Currently, most methods focus on more efficient data sampling. Our approach shows a significant speedup by combining measurement speedup and a reduction in the number of measurements needed to tune a quantum dot device. Using an autotuning algorithm based on a neural network and faster measurements by harnessing the FPGA embedded in Keysight’s Quantum Engineering Toolkit (QET), the measurement time of stability diagrams has been reduced by a factor of 9.8. This led to an acceleration factor of 2.2 for the total initialization time of a SiGe quantum dot into the single-electron regime, which is limited by the Python code execution.
\end{abstract}


\maketitle

\section{Introduction}
\label{sec:introduction}
Spin qubits are a promising quantum computing architecture thanks to their long coherence time\cite{Tyryshkin2011,Muhonen2014,Saeedi2013,Veldhorst2014} and compatibility with industrial semiconductor manufacturing\cite{Elsayed2024,Bartee2024,Veldhorst2017,urdampilleta_gate-based_2019}. As a first tuning step to operate multidot qubit systems, stability diagram measurements are essential to initialize a quantum dot in the desired charge state and maintain that charge state despite drift. However, for large systems with hundreds or even millions of qubits required for useful quantum applications, this tuning becomes challenging and time consuming. To speed up this tuning process, many approaches have been proposed and are described below.

For instance, more efficient sampling enables the extraction of the desired charge state information from stability diagrams without the need to explore the entire voltage space \cite{Lennon2019, Zwolak2021}. However, sparse measurements work best if the measurement time is much longer than the time needed by the algorithm to choose the next measurement, which is not always the case for RF measurement schemes. 

To be more efficient, an algorithm can also be used to automatically tune a device while also using sparse measurements. Many autotuning procedures have been proposed to remove the need for human intervention in the tuning process, one of which even required less time than human experts\cite{Moon2020}. Regardless, more work still has to be done to develop a general autotuning algorithm that can adapt to different types of devices and variability in the fabrication process\cite{Zwolak2023}. Many algorithms typically focus on one of the steps in the tuning process \cite{Zwolak2023} such as bootstrapping to estimate voltage ranges for operation and tune charge sensors\cite{Kovach2024}, coarse tuning to form the desired number of quantum dots, establishing controllability to remove capacitive crosstalk between gates with virtual gates\cite{Mills2019, rao_mavis_2025}, charge-state tuning to adjust the number of charges in each quantum dot\cite{Lapointe-Major2020, Czischek2021, Ziegler2023, Durrer2020, Zwolak2020, yon_experimental_2025} and fine-tuning to perform logical operations on the qubits\cite{VanEsbroeck2020, Schuff2023}. On the other hand, in some publications, steps from bootstrapping to either coarse tuning \cite{Moon2020, Zubchenko2025, Severin2024}, charge-state tuning\cite{baart_computer-automated_2016} or even fine-tuning\cite{Schuff2024} are automated using their respective algorithm. Steps from coarse tuning to establishing controllability and charge-state tuning have also been automated \cite{Liu2022}.  

Still, even if some autotuning algorithms achieved faster tuning than the manual approach, DC stability diagram measurements were still identified as a limiting factor in many cases \cite{Moon2020, Zwolak2020, VanEsbroeck2020, Severin2024, baart_computer-automated_2016, Schuff2024, yon_experimental_2025}.

Fast voltage sweeps, such as video-mode measurements \cite{Stehlik2015}, are typically done with an arbitrary waveform generator (AWG) where a predefined waveform is loaded into memory and then generated after a trigger is received. Therefore, the amount of on-board memory limits the voltage window that can be measured. It is also not possible to rapidly measure voltage regions that are far apart, since slow DC voltages must be adjusted to move the fast measurement window. 

Qubit measurement platforms with on-board field-programmable gate arrays (FPGAs) appear as a potential solution, as they allow more flexibility and fast tuning for feedback during measurements. In-house solutions were developed to get the flexibility and performance of FPGAs at an affordable price\cite{Xu2021, Stefanazzi2022, Toubeix2024, Park2022}. Still, they require much more programming than commercial solutions.

Companies such as Keysight Technologies \cite{Nakajima2021,Philips2022,stemp_tomography_2024}, Quantum Machines \cite{Berritta2024, park_passive_2025}, Qblox \cite{de_smet_high-fidelity_2025} and Zurich Instruments \cite{xue_sisige_2024, park_passive_2025} each offer a platform with the option to program experiment sequences in an FPGA, using a user-friendly interface, to achieve precise control over the generated signals. We chose Keysight’s Quantum Engineering Toolkit (QET) because it has an open “sandbox” that allows adding custom FPGA code to the instrument, expanding its capabilities beyond those of the included libraries.

Therefore, this work aims to speed up quantum dot charge-state tuning by using the Quantum Engineering Toolkit’s ability to program the stability diagram measurement sequence inside its FPGAs alongside an efficient autotuning algorithm. Furthermore, we take advantage of the fact that the autotuning algorithm can freely explore the voltage space at minimal cost in terms of time thanks to the low latency of the FPGA. This enables the use of more complex search patterns to increase the tuning success rate while still rapidly reaching the single-electron regime. This is an important step towards the tuning of larger arrays of quantum dots used for quantum computing, for which fast and reliable tuning without human intervention is essential.

\section{Methods}
\subsection{Device characterization}
\begin{figure}[t!]
\includegraphics[width=\columnwidth]{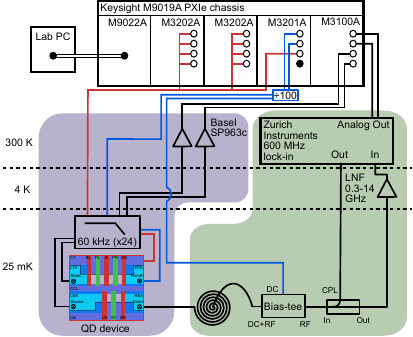}
\caption{\justifying Experimental setup with the DC circuit on the left in purple and the reflectometry circuit on the right in green. The Keysight chassis handles control and readout of the SiGe quantum dot device. On the quantum dot (QD) device schematic, red gates act as barrier gates, green gates as plunger gates, blue gates as reservoirs and purple gates as confinement gates. A voltage divider is used to increase the resolution of the M3201A's voltage applied to the ohmic contacts by a factor of 100.}
\label{setup}
\end{figure}

Experiments are performed on a SiGe quantum dot (QD) device fabricated in a 300 mm platform at IMEC. Information on the design and device fabrication can be found here \cite{koch_industrial_2025}. We operate the device with a single quantum dot whose charge state is measured with a single-electron transistor (SET). A schematic of the experimental setup is shown in Fig.~\ref{setup}.

DC voltages are applied on the gates through lines with a 60 kHz low-pass filter using M3202A and M3201A arbitrary waveform generators (AWGs) from the Keysight chassis, while a M3100A digitizer housed in the same chassis is used for signal acquisition. Current measurements are done through the ohmics, using I/V converters for amplification, 
to check for possible leaks in the sample.

Reflectometry measurements are performed using an RF-SET configuration \cite{Schoelkopf1998, zirkle_radio_2020, rivard_multimode_2025} to measure stability diagrams, enable faster readout and obtain a better signal-to-noise ratio using a Zurich Instruments UHFLI, chosen for its 600 MHz bandwidth. The lock-in amplifier is used to send RF excitations to the SET ohmic through a 225 MHz LC resonant circuit composed of a superconducting Nb spiral inductor. The amplitude and phase signals are then sent to the Keysight digitizer using the analog outputs so that the whole experiment can be controlled by the Keysight chassis. 

The stability diagram of the dot, shown in Fig.~\ref{diagrams}a, defines the voltage space that the autotuning algorithm can explore. Using an FPGA sequence to measure a stability diagram removes the communication time per point, which speeds up the acquisition, especially for fast measurement schemes like reflectometry. Another advantage of the on-the-fly generated waveforms is that there is no memory limit for waveforms, as it is usually the case with AWGs. Therefore, large stability diagrams with hundreds of points and spanning several hundreds of millivolts per axis, like in Fig.~\ref{diagrams}, can be measured faster without running into memory limitations. To give an order of magnitude, by simply removing the communication latency without changing any other parameter, this stability diagram can be measured approximately 5x faster.

\subsection{Autotuning and FPGA Measurement Software}
PathWave Test Sync Executive \cite{PWTSE} is used to program the measurement sequence for the stability diagrams in the FPGAs of the AWGs and digitizers. This program facilitates the creation of FPGA sequences using Python, a programming language that is more accessible to physicists than VHDL. The program generates two waveforms on the fly, one for each axis of the stability diagram, to set the gate voltages using any two channels from the AWGs in the chassis and then up to four channels of the digitizer can be measured simultaneously for readout. All signal synchronization between the AWG channels and the digitizer is handled by PathWave Test Sync Executive, which utilizes the chassis' backplane. The FPGA source code is available on GitHub \cite{roux_hvi-sweeper_2025}. 

\begin{figure}[t!]
\includegraphics[width=\columnwidth]{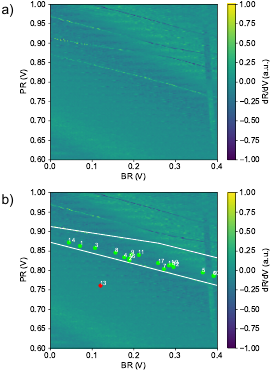}
\caption{\justifying a) Single quantum dot stability diagram used as the voltage space for the autotuning algorithm to explore. The diagram has four visible transition lines, but the two lines in the middle change slope and have an anticrossing at (0.22, 0.9) V, indicating the presence of a parasitic dot coupling to the quantum dot under investigation. b) Result of 19 reflectometry autotuning runs with an integration time of 16.6~ms and a slew rate of 10~V/s. The autotuning runs that converge in the single-electron regime (surrounded by white lines) are marked in green. In this case, the autotuning success rate is 18/19, or 95\%.}
\label{diagrams}
\end{figure}

Autotuning to identify the single-electron regime is performed using the exploration algorithm described in \cite{Yon2024} where a convolutional neural network is used to detect the charge transition lines on small 18~mV x 18~mV voltage patches with 1~mV pixels. Thanks to the fast voltage ramps, the algorithm can explore the voltage space in many directions with an X-shaped pattern to cover a large area with only a minimal time penalty in order to find the first transition. When an autotuning run finishes in the single-electron regime, marked with white lines in Fig.~\ref{diagrams}b, the run is considered a success, and the voltage coordinates given at the output of the autotuning algorithm are marked in green. Otherwise, the coordinates are marked in red. For the example in Fig.~\ref{diagrams}b, the success rate is 95\%. To get statistics of the success rate of the autotuning procedure (ratio of successes over failures), a set of 19 autotuning runs with random starting points measured by reflectometry is done for each integration time and slew rate. To facilitate the comparison of autotuning times for different experimental conditions, the starting points are set using a pseudo-random number generator to ensure that autotuning runs with the same number have the same starting point. 

To quantify the performance of the autotuning procedure, the duration of each step is measured independently. The total autotuning time is the sum of the time used by the tuning algorithm for decision-making, the measurement time, and the time required to save the results of the autotuning run. The decision-making process includes plotting and saving stability diagram patches, detecting lines using the convolutional neural network with PyTorch, and selecting the next patch to measure, in Python, based on the CNN output. The measurement time includes the initialization of the measurement and data acquisition. Finally, the data saved for each autotuning run includes final voltage coordinates, timers, and a video of the full tuning procedure for review if needed.

\section{Results}

\begin{figure}[t!]
\includegraphics[width=\columnwidth]{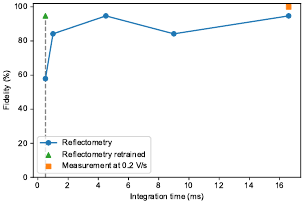}
\caption{\justifying Success rate of the autotuning procedure as a function of integration time. A grey line is added at 0.5~ms, where the autotuning measurement has a lower success rate. The green triangle  represents the same reflectometry measurements done at 0.5~ms of integration time but with the neural network retrained with noisy reflectometry data. The orange square represents the measurement taken at 16.6~ms, but with a lower 0.2~V/s slew rate.}
\label{fidelity_time}
\end{figure}

To determine whether the success rate remains high for shorter integration times, additional autotuning runs are performed at different integration times, and the results are summarized in Fig.~\ref{fidelity_time}. It shows the average success rate for sets of 19 reflectometry measurements at different integration times. The success rate hovers around 90\%, except for 0.5~ms of integration time, where it drops to 58\%, due to increased noise. This shows, however, that the line detection efficiency of the model is robust to some extent, since the model was originally trained only on data acquired by current measurements and not reflectometry. We further demonstrate that retraining of the neural network with noisier reflectometry data (acquired with 0.5~ms integration time) increases this success rate from 58\% back to 95\%. As no change in the autotuning speed is noticed when reducing the integration time from 1~ms to 0.5~ms, lower integration times are not considered.

We then investigate the effect of the slew rate on the autotuning success rate and compare the success rates obtained for slew rates of 0.2~V/s and 10~V/s. For an integration time of 16.6 ms, reducing the slew rate from 10 V/s to 0.2 V/s increases the autotuning success rate from 95\% to 100\%. However, for integration times of 16.6~ms or lower, the sweep time of a 0.2 V/s slew rate would account for more than a quarter of the measurement time and become a bottleneck. Thus, the marginal improvement in success rate does not justify the increased autotuning time. 

\begin{figure}[t!]
\includegraphics[width=\columnwidth]{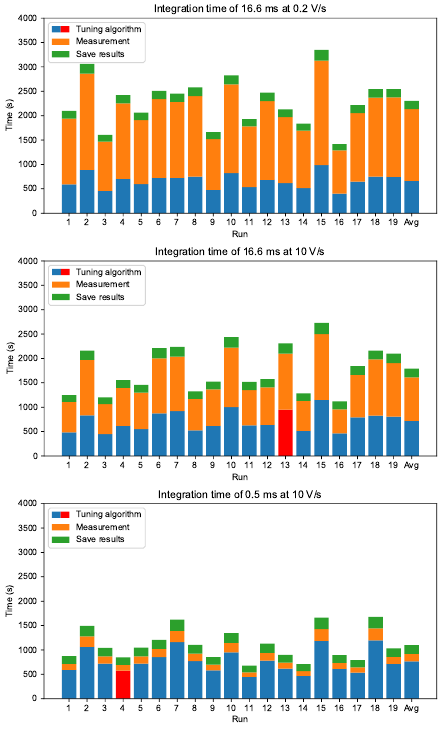}
\caption{\justifying Bar plot showing the time needed for each autotuning run to complete. The total time is displayed with bars and is divided by the decision-making time between measurements (in blue for successful runs and red for failed runs), the measurement time (orange), and the time needed to save results (green). The measurements were made using reflectometry with an integration time of 16.6~ms at 0.2~V/s, 16.6~ms at 10~V/s and 0.5~ms at 10~V/s from top to bottom. The last bar of each plot represents the average of the 19 runs in the plot.}
\label{histograms}
\end{figure}

\begin{figure}[t!]
\includegraphics[width=\columnwidth]{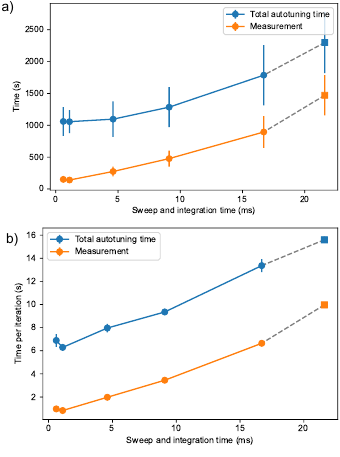}
\caption{\justifying Total autotuning time (blue) and measurement time (orange) as a function of sweep and integration time are shown at the top. Same figure at the bottom but the data is normalized by the number of steps taken by the autotuning algorithm. Data is averaged over the 19 runs for each point and the standard deviation is shown as error bars. The sweep and integration time is the sum of the integration time, and the time, per point, needed to sweep to the target voltage. Therefore, it takes into account the slew rate used for the measurements. The square data point was acquired using an integration time of 16.6~ms and a slew rate of 0.2~V/s while the other points were acquired using a slew rate of 10~V/s which makes the sweep time negligible.}
\label{autotuning_time}
\end{figure}

In order to assess the bottlenecks going from an integration time of 16.6 ms to 0.5 ms, the duration of each run for these integration times is shown in Fig.~\ref{histograms}. The figure shows bar plots of the autotuning time for all 19 autotuning runs for three integration times: 16.6~ms at 0.2~V/s, 16.6~ms at 10~V/s and 0.5~ms at 10~V/s, from top to bottom. Variations in autotuning time are mostly due to the random starting point of each run, which influences the number of iterations the algorithm needs to reach the single-electron regime. When normalized by the number of steps used by the autotuning algorithm, the total autotuning time remains fairly constant from run to run (see Fig.~\ref{autotuning_time}). The average measurement time per iteration is (10.0 $\pm$ 0.2) seconds at 16.6~ms integration with a 0.2~V/s slew rate, which accounts for 64\% of the total autotuning time.With the same integration time but a 10~V/s slew rate, the average measurement time per iteration is (6.65 $\pm$ 0.03) seconds, which represents 50\% of the total autotuning time. At an integration time of 0.5~ms and a 10~V/s slew rate, the average measurement time remains fairly constant at around (0.95 $\pm$ 0.03) seconds per iteration. This represents only 14\% of the total autotuning time. These numbers show that the slew rate and integration time have a significant impact on the autotuning time and also indirectly highlight the impact of communication latency on measurement speed, since latency is effectively limiting the slew rate (see appendix). To estimate the fastest possible autotuning time when measurements account for only 14\% of the total time, the average autotuning time is calculated for each integration time, and the result is shown in Fig.~\ref{autotuning_time}.

Fig.~\ref{autotuning_time}a shows the total autotuning time and measurement time averaged over the 19 random starting points. In Fig.~\ref{autotuning_time}b, the same data is normalized by the number of iterations used by the algorithm before averaging over the 19 runs. The standard deviation is shown as error bars, which are much bigger in Fig.~\ref{autotuning_time}a since the total autotuning and measurement times depend on the number of iterations needed to reach the single-electron regime. Those times are almost constant when divided by the number of iterations. When comparing the total autotuning time from the longest to the shortest sweep and integration time, going from 2301 s to 1062 s, a reduction of 2.2x is obtained. When considering only the measurement time, going from 1471 s to 150 s, a reduction of 9.8x is obtained. Decreasing the integration time to 0.5 ms does not significantly decrease the total autotuning time. Indeed, as it can be seen in Fig.~\ref{histograms}, at an integration time of 0.5 ms, the decision-making of the algorithm takes 70\% of the total autotuning time. The measurement time tends towards a constant value, despite decreasing integration time. This behavior may be attributed to the initialization time of the measurement. Consequently, improving the line detection code and measurement initialization, both written in Python, could reduce the overall autotuning time. Also, some operations could be parallelized in Python or the plotting and saving of the stability diagram patches could be disabled. 

\section{Conclusion}
In summary, we used an FPGA to control stability diagram measurements, reducing communication latency and fully utilizing shorter integration times and faster slew rates for faster stability diagram measurements and faster voltage space exploration. Using this approach, we were able to automatically tune, with a success rate of 90\%, a SiGe quantum dot into the single-electron regime using a neural network for charge transition line detection and Keysight’s Quantum Engineering Toolkit for the FPGA-accelerated measurements. Compared to a measurement limited by communication latency, autotuning was done 2.2x faster, while the measurements were done 9.8x faster. This represents an important step towards accelerating quantum dot tuning, since the low latency of FPGAs remains underutilized in the community. To build on this work, the decision-making process of the tuning algorithm and the neural network could be integrated into the instrument’s FPGAs. This would significantly reduce the decision time between measurements.

\section{Appendix}\label{appendix}
\subsection*{Slew rate limited by communication}
To obtain a significant speedup in measurement time, communication between the instruments and the lab computer must be removed. When using an oscilloscope to measure the average time needed to step the AWG voltage and read the digitizer value with a negligible integration time of 1.5~\textmu s, using the lab computer instead of the FPGA, the latency is estimated to be around 27~ms. Therefore, taking advantage of the FPGA inside the instruments to control the experiment and removing this latency is crucial to obtain any significant speedup in measurement time. Some instruments can also generate voltage ramps internally without requiring communication at every voltage step, which also removes the majority of the communication latency. Without either of those approaches, the added latency translates into a slower effective slew rate. For example, taking into account the latency of 27~ms and using the same integration time of 1.5~\textmu s, the maximum slew rate possible is 0.39~V/s using 10~mV steps or 0.04~V/s using 1~mV steps. Therefore, the measurements with a slew rate of 0.2~V/s and an integration time of 16.6~ms shown in Fig.~\ref{autotuning_time} represents a measurement limited by communication latency.

\section*{Acknowledgment}
The authors would like to thank Philip Krantz from Keysight Technologies for facilitating our collaboration with their technical team. We acknowledge the support of the Natural Sciences and Engineering Research Council of Canada (NSERC), [funding reference number RGPIN–2020–0573, 515827-2017 and 546399-2019] and the Fonds de recherche du Québec (\url{https://doi.org/10.69777/268397}). 
We also acknowledge the support of the Canada Foundation for Innovation (Atomistics of Engineering and Natural Materials) and of the National Research Council Canada (Quantum Sensors Challenge Program 20-1).
This research was also undertaken thanks in part to funding from the Canada First Research Excellence Fund and the NSERC‐CREATE program QSciTech.

\bibliography{biblio.bib}
\bibliographystyle{IEEEtran}

\end{document}